\documentclass[aps,showpacs,prd,twocolumn]{revtex4}
\usepackage{amssymb}
\usepackage{amsfonts}
\usepackage{amsmath}
\usepackage{graphicx}
\usepackage{color}
\usepackage[dvips]{epsfig}
\usepackage[dvips]{graphicx}
\usepackage{float}

\begin{document}

\title{Maxwell-Higgs self-dual solitons on an infinite cylinder}
\author{Rodolfo Casana}
\email{rodolfo.casana@ufma.br}
\affiliation{Departamento de F\'{\i}sica, Universidade Federal do Maranh\~{a}o,
65080-805, S\~{a}o Lu\'{\i}s, Maranh\~{a}o, Brazil}
\author{Lucas Sourrouille}
\email{sourrou@df.uba.ar}
\affiliation{Universidad Nacional Arturo Jauretche, 1888, Florencio Varela, Buenos Aires,
Argentina}

\begin{abstract}
We have studied the Maxwell-Higgs model on the surface of an infinite
cylinder. In particular we show that this model supports self-dual
topological soliton solutions on the infinite tube. Finally, the
Bogomol'nyi-type equations are studied from theoretical and numerical point
of view.
\end{abstract}

\maketitle



It is well know, that the abelian Higgs models in (2+1)-dimension with the
Maxwell term present topological stable vortex solution \cite{NyO} (for
review see \cite{hor, hor1, hor2, hor3}). These models have the particularity to
became auto-dual when the self-interactions are suitably chosen. When this
occur the model presents particular mathematical and physics properties,
such as the supersymmetric extension of the model\cite{WO}, and the
reduction of the motion equation to first order derivative equation\cite%
{Bogo, Bogo1}. Characteristically, these vortex solutions carry magnetic flux but
are electrically neutral. They correspond to vortex-like object carrying a
magnetic field in its core, where the scalar field vanishes. Also, the
magnetic flux is quantized.

Recently, there has been interest in the study of gauge theories on infinite
tubes \cite{thooft1}. Particularly, it was studied $SU(2)$ Yang-Mills model
within closed and open tubes, showing the existence of magnetic monopoles
and dyons living within this tube-shaped domain.

Here we are interested on the study of the abelian Maxwell-Higgs model
within an infinite cylinder. Our goal is to show that the abelian
Maxwell-Higgs model support topological solitons when is defined within an
infinite cylinder. Specifically, we propose an ansatz such that the model is
reduced to a theory within an infinite cylinder surface. We will show that
for the model defined in the cylinder it is possible to obtain self-dual or
Bogomolnyi equations by minimizing the energy functional of the model. The
model is then bounded below by topological number. Finally we analyze the
numerical solutions of the field equations showing the behavior of matter
and gauge fields as well as their energy density. We also analyze the
behavior of the magnetic vortex.

We start by considering the Maxwell-Higgs electrodynamics in a
(2+1)-dimensional curved manifold described by the following metric%
\begin{equation}
ds^{2}=dt^{2}- \rho^2 d\phi ^{2}-dz^{2},
\end{equation}%
(for solitons on curved manifold see \cite{hor0, hor01, hor02, hor03})
where the azimuthal angle runs over $0\leq \phi \leq 2\pi $
,and the $z$-coordinate varies between $0\leq
z_+<+\infty$. Here, $\rho$ is a constant and in the following
we set $\rho=1$, for simplicity.
Such a manifold is a
semi-infinite cylinder immersed in a (3+1)-dimensional
space-time. The action describing the Maxwell electrodynamics coupled to
complex scalar field $\psi (x)$ is 
\begin{equation}
S=\int d^{3}x\Big[-\frac{1}{4}F_{\mu \nu }F^{\mu \nu }+|D_{\mu }\psi
|^{2}-U(|\psi |)\Big],  \label{Ac7}
\end{equation}%
The covariant derivative is defined as $D_{\mu }=\partial _{\mu }+ieA_{\mu }$
and the metric tensor is $g^{\mu \nu }=(1,-1,-1)$.

By varying with respect to $A_{\mu }$ and $\psi ^{\dagger }$, we obtain the
field equations 
\begin{equation}
\partial ^{\mu }F_{\mu \nu }+J_{\nu }=0\;,  \label{EqM5}
\end{equation}%
\begin{equation}
D_{\mu }D_{\mu }\psi +\frac{\partial U}{\partial \psi ^{\ast }}=0\;,
\label{EqM4}
\end{equation}%
where $J_{\nu }=-ie\Big[\psi ^{\ast }(D_{\nu }\psi )-\psi (D_{\nu }\psi
)^{\ast }\Big]$.

The energy associated with the action (\ref{Ac7}) is 
\begin{equation}
E=\int d^{3}x\Big[\frac{1}{2}F_{0i}^{2}+\frac{1}{4}F_{ji}^{2}+|D_{0}\psi
|^{2}+|D_{i}\psi |^{2}+U(|\psi |)\Big].  \label{Ac8}
\end{equation}

From (\ref{EqM5}), the stationary Gauss law reads 
\begin{equation}
\partial _{j}\partial _{j}A_{0}+2e^{2}A_{0}|\psi| ^{2}=0~,  \label{rc1}
\end{equation}%
it is satisfy identically by the configuration $A_{0}=0$, thus, in the
static regimen the model describes pure magnetic configurations.

In this note we are interested in the particular case of the static field
configurations with $A_{0}=0$. In this context, the energy in (\ref{Ac8})
reduces to: 
\begin{equation}
E=\int d^{2}x\Big[\frac{1}{2}\mathbf{B}^{2}+|D_{i}\psi |^{2}+U(|\psi |)\Big],
\label{Ac9}
\end{equation}%
where $\mathbf{B}=\nabla \times \mathbf{A}$. 
\\
Let us consider the following ansatz: 
\begin{equation}
A_{z_{+}}=0\,,\;\;A_{\phi }=\frac{h(z_{+})}{e}\,,\;\;\psi
=vf(z_{+})e^{-in\phi }\,,  \label{ansatz}
\end{equation}%
where $n$ is a topological number called the winding number. With such
conditions we see that the magnetic field restricted to the cylinder surface
reduce to: 
\begin{equation}
\mathbf{B}=\nabla \times \mathbf{A}=-\partial _{z_{+}}A_{\phi }\mathbf{\hat{%
\rho}}=-\frac{\partial _{z_{+}}h(z_{+})}{e}\mathbf{\hat{\rho}},  \label{magr}
\end{equation}%
In addition the $\phi $-component of the covariant derivative takes the
form, 
\begin{equation}
D_{\phi }\psi =\partial _{\phi }\psi +ieA_{\phi }\psi ,
\end{equation}%
where we have used $\rho =1$. Using the ansatz (\ref{ansatz}) we can replace 
$\psi $ and $A_{\phi }$, so that, 
\begin{eqnarray}
D_{\phi }\psi  &=&-invf(z_{+})e^{-in\phi }+ivf(z_{+})h(z_{+})e^{-in\phi } 
\notag \\[0.15cm]
&=&ivf(z_{+})\left[ -n+h(z_{+})\right] e^{-in\phi }.
\end{eqnarray}%
Thus, we may develop the term $|D_{\phi }\psi |^{2}$ to be, 
\begin{equation}
|D_{\phi }\psi |^{2}=v^{2}f(z_{+})^{2}[h(z_{+})-n]^{2}\;,
\end{equation}%
On the other hand, the component $\mathbf{\hat{z_{+}}}$ of the covariant
derivative reduce to 
\begin{equation}
D_{z_{+}}\psi =\partial _{z_{+}}\psi =ve^{-in\phi }\partial
_{z_{+}}f(z_{+})\;,
\end{equation}%
so that, 
\begin{equation}
|D_{z_{+}}\phi |^{2}=v^{2}[\partial _{z_{+}}f(z_{+})]^{2}.
\end{equation}%
Thus, it is possible, in order to establish the suitable boundary
conditions, rewrite the energy (\ref{Ac9}) in terms of the ansatz (\ref%
{ansatz}) 
\begin{eqnarray}
E &=&\int dz_{+}d\phi \left[ \frac{1}{2}\left( \frac{\partial
_{z_{+}}h(z_{+})}{e}\right) ^{2}+U\left( f\right) \right.   \label{Ac10} \\
&&\hspace{0cm}\left. \frac{{}}{{}}+v^{2}\left[ \partial _{z_{+}}f(z_{+})%
\right] ^{2}+v^{2}f(z_{+})^{2}\left[ h(z_{+})-n\right] ^{2}\right] ,  \notag
\end{eqnarray}%
The field equations (\ref{EqM5}) and (\ref{EqM4}) may also be written in
terms of the ansatz. The set of equations (\ref{EqM5}) reduce to one
equation: 
\begin{equation}
\partial _{z_{+}}^{2}h(z_{+})=2v^{2}e^{2}[f(z_{+})]^{2}[h(z_{+})-n]\;,
\label{EqM6}
\end{equation}%
whereas the equation (\ref{EqM4}) reads as: 
\begin{equation}
\partial _{z_{+}}^{2}f(z_{+})-f(z_{+})[h(z_{+})-n]^{2}-\frac{1}{2v}\frac{%
\partial U}{\partial f}=0\;.  \label{EqM7}
\end{equation}%
Appropriate boundary conditions for a soliton solution of finite energy
should take the form 
\begin{equation}
f(+\infty )=1\,,\;\;\;\;\;\ f(0)=\gamma _{0}^{\left( n\right) },  \label{b}
\end{equation}%
\begin{equation}
h(+\infty )=n\,,\;\;\;\;\;\ h(0)=0,  \label{b1}
\end{equation}%
where $\gamma _{0}^{\left( n\right) }$ is related to the winding number and therefore is uniquely
determined for each topological sector. In particular we will show by numerical analysis that, $0<\gamma 
_{0}^{\left( n\right) }<1$. Note, that this
requirement, does not affect the regularity of the field $\psi (\phi, z)$ at $%
z_{+}=0$, since $z_{+}=0$ is a circle $S_{1}$ and then $\psi (\phi ,0)$ is a
map $S_{1}\rightarrow S_{1}$, i.e. for each angle $\phi $ we have a complex
number $v\gamma^{(n)}_{0}e^{-in\phi }$.

The boundary conditions (\ref{b1}) allow to compute the magnetic
flux, the important observable for the solitons described by the
Maxwell-Higgs electrodynamics. So, by using the formulas (\ref{magr})
and (\ref{b1}), the magnetic flux reads %
\begin{eqnarray}
\Phi &=&\int dz_{+}d\phi ~B_{\rho }  \notag \\
&=&2\pi \int_{0}^{\infty }dz_{+}\left( -\frac{\partial _{z_{+}}h(z_{+})}{e}%
\right) =-\frac{2\pi }{e}n,  \label{flux}
\end{eqnarray}%
as expected, it becomes a quantized quantity, i.e., proportional to
the winding number $n$ describing the respective topological sector.%

In order to find the static field configurations that are stationary points
of the energy, it is convenient to rewrite the expression of the energy as 
\begin{eqnarray}
E &=&2\pi \!\int_{0}^{\infty }\!dz_{+}\!\left[ \frac{1}{2}\left( \frac{%
\partial _{z_{+}}h(z_{+})}{e}\pm \sqrt{2U}\right) ^{2}\right.   \notag \\%
[0.15cm]
&&\hspace{-0.5cm}+v^{2}\left( \frac{{}}{{}}\partial _{z_{+}}f(z_{+})\pm
f(z_{+})\left[ h(z_{+})-n\right] \right) ^{2}  \label{Ac11} \\[0.15cm]
&&\hspace{-0.5cm}\left. \mp \frac{\partial _{z_{+}}h(z_{+})}{e}\sqrt{2U}\mp
2v^{2}[h(z_{+})-n]f(z_{+})\partial _{z_{+}}f(z_{+})\right] .  \notag
\end{eqnarray}%
Let us now define the the potential term $U(f)$ to be the usual symmetry
breaking one of the Maxwell-Higgs model in its self-dual form, 
\begin{equation}
U\left( f\right) =\frac{e^{2}v^{4}}{2}\left( f^{2}-1\right) ^{2}.
\label{pott}
\end{equation}%
Now it is not difficult to see that%
\begin{equation}
\partial _{z_{+}}\left( \frac{\sqrt{2U}}{e}\right) =2v^{2}f(z_{+})\partial
_{z_{+}}f(z_{+}),
\end{equation}%
which allow to rewrite the third line of Eq. (\ref{Ac11}) in the
following way
\begin{eqnarray}
&&\mp \frac{\partial _{z_{+}}h(z_{+})}{e}\sqrt{2U}\mp
2v^{2}[h(z_{+})-n]f(z_{+})\partial _{z_{+}}f(z_{+})  \notag \\
&=&\mp \frac{1}{e}\partial _{z_{+}}\left( \left[ h(z_{+})-n\right] \sqrt{2U}%
\right) .
\end{eqnarray}%
Therefore, the energy (\ref{Ac11}) may be rewritten as a sum of
squared terms plus a total derivative, 
\begin{eqnarray}
E &=&\!2\pi \int_{0}^{\infty }\!\!dz_{+}\left[ \frac{1}{2}\left( \frac{%
\partial _{z_{+}}h(z_{+})}{e}\pm \sqrt{2U}\right) ^{2}\right.   \notag \\%
[0.15cm]
&&\hspace{0.5cm}+v^{2}\left( \frac{{}}{{}}\partial _{z_{+}}f(z_{+})\pm
f(z_{+})\left[ h(z_{+})-n\right] \right) ^{2}  \notag \\[0.15cm]
&&\hspace{0.5cm}\left. \frac{{}}{{}}\mp \frac{1}{e}\partial _{z_{+}}\left(
(h(z_{+})-n)\sqrt{2U}\right) \right] .
\end{eqnarray}%
The total derivative may be explicitly evaluated by using the expression (%
\ref{pott}) and the boundary conditions (\ref{b}) and (\ref{b1}). In such
case we have, 
\begin{eqnarray}
&&\mp 2\pi v^{2}\int_{0}^{\infty }dz_{+}\ \partial _{z_{+}}\left[ \left(
h(z_{+})-n\right) \left( f\left( z_{+}\right) ^{2}-1\right) \right]   \notag
\\
&=&\pm 2\pi v^{2}n\left[ 1-\left( \gamma _{0}^{\left( n\right) }\right) ^{2}%
\right] 
\label{bound}
\end{eqnarray}%
Therefore, we see that the energy is bounded below by a multiple of the
magnitude of the winding number (for positive $n[1-\gamma _{0}^{2}]$ we
choose the upper sign, and for negative $n[1-\gamma _{0}^{2}]$ we choose the
lower sign).
Here, it is interesting to note that the topological bound (\ref{bound}) consist on two terms.
The first of these terms is the topological bound    
of the energy associated to the $(2+1)$ planar Maxwell-Higgs model, i.e. 
\begin{equation}
 \pm 2\pi v^2 n\;,
 \label{bound1}
\end{equation}
which is proportional to the magnetic flux (\ref{flux}).
The second term is 
\begin{equation}
\mp 2\pi v^{2}n \left( \gamma _{0}^{\left( n\right) }\right) ^{2}
 \label{bound2}
\end{equation} 
and it is a novel term that emerges in our theory.
As we mention $\gamma _{0}^{\left( n\right) }$ depends on the winding number $n$, 
so that the topological bound (\ref{bound}) is unequivocally determined for each topological sector.
In particular we will show, by numerical analysis, that $\gamma
_{0}^{\left( n\right) }\rightarrow 0$ for large values of $n$, such that 
\begin{eqnarray}
\pm 2\pi v^{2}n \Big[ 1- \Big( \gamma _{0}^{( n) }\Big) ^{2}\Big]\rightarrow \pm 2\pi v^2 n
\end{eqnarray}
Thus, our topological bound tends, for large values of $n$, to the topological bound (\ref{bound1})
of the Maxwell-Higgs model. 
\\
The bound (\ref{bound}) is saturated by fields satisfying the first-order
Bogomol'nyi self-duality equations, 
\begin{equation}
\frac{\partial _{z_{+}}h(z_{+})}{e}\pm ev^{2}\left( f^{2}-1\right) =0
\label{bogo01}
\end{equation}%
\begin{equation}
\partial _{z_{+}}f(z_{+})\pm f(z_{+})(h(z_{+})-n)=0  \label{bogo1}
\end{equation}%
It easily verified that BPS\ equations (\ref{bogo1}) and (\ref{bogo01})
solve the second-order equations of motion given in (\ref{EqM6})\ and (\ref%
{EqM7}) if the potential is given by (\ref{pott}).

On the other hand, by using self dual equations, the BPS energy density $%
\varepsilon \left( z_+\right) $, $\left[ E=2\pi \rho _{0}\int_{0}^{\infty
}dz_+~\varepsilon \left( z_+\right) \right] $, is expressed as 
\begin{equation}
\varepsilon \left( z_+\right) =e^{2}v^{4}\left( f^{2}-1\right) ^{2}+
2v^{2}f^{2}\left( h-n\right) ^{2},
\end{equation}%
which is a positive-definite quantity.

Let us now concentrate in the solutions of the equations (\ref%
{bogo1}) and (\ref{bogo01}) at $z_+\rightarrow +\infty $. From the preceding
arguments it is not difficult to find their asymptotic behavior 
\begin{eqnarray}
f(z_+)&\simeq&1-\lambda _{1}e^{-ev\sqrt{2}z_+}  \notag \\[-0.1cm]
\\[-0.1cm]
h(z_+)&\simeq&n-\lambda _{1}ev\sqrt{2}e^{-ev\sqrt{2}z_+}\;,  \notag
\end{eqnarray}
with $\lambda _{1}$ a real number. From these solutions we may evaluate the
magnetic field at $z_+\rightarrow +\infty$. By substituting $h(z_+)$ in the
expression (\ref{magr}) for the magnetic field we have 
\begin{equation}
{B}|_{z_+\rightarrow \infty }=-2ev^{2}\lambda _{1}e^{-ev\sqrt{2}z_+}\;,
\label{magr1}
\end{equation}%
which shows the asymptotic behavior of the magnetic field.

We also analyze the behavior of the solution at $z_{+}\rightarrow 0$. In
such case we have for $n>0:$ 
\begin{eqnarray}
f(z_{+}) &=&\gamma _{0}^{\left( n\right) }-n\gamma _{0}^{\left( n\right)
}z_{+}+...,  \notag  \label{bogo12} \\[-0.15cm]
&& \\[-0.15cm]
h(z_{+}) &=&e^{2}v^{2}\left[ 1-\left( \gamma _{0}^{\left( n\right) }\right)
^{2}\right] z_{+}-e^{2}v^{2}\left( \gamma _{0}^{\left( n\right) }\right)
^{2}nz_{+}^{2}+...,  \notag
\end{eqnarray}%
where $\gamma _{0}$ is a real number depending on the value of $n$ and it is
determined numerically. Finally, we can evaluate the magnetic field near $%
z_{+}=0$ by substituting the last equation of (\ref{bogo12}) in the
expression (\ref{magr}), 
\begin{equation}
{B}|_{z_{+}\rightarrow 0}=-ev^{2}\left[ 1-\left( \gamma _{0}^{\left(
n\right) }\right) ^{2}\right] \;.  \label{magr2}
\end{equation}%
By comparing the expression (\ref{magr2}) with (\ref{magr1}) we see that the
absolute value of the magnetic field has a maximum at $z_{+}=0$ and decrease
exponentially as $z_{+}\rightarrow +\infty $.

We have verified that the self dual equations solved with boundary
conditions (\ref{b}) and (\ref{b1}) provide well-behaved solutions both at
origin as at infinity.

Below we show the numerical solutions of the self-dual equations ( \ref%
{bogo01}) and (\ref{bogo1}) with boundary conditions given by (\ref{b}) and (%
\ref{b1}) for some values of $n>0$. To perform the numerical analysis we
have considered $e=v=\rho _{0}=1$ and $n=1,2,3,4,5$.

\begin{figure}[tbp]
\centering\includegraphics[width=8.25cm]{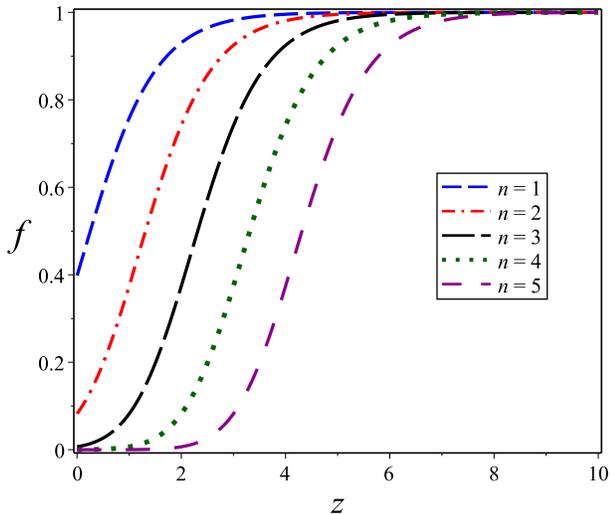}
\caption{The scalar field profile $f(z_{+})$.}
\label{f_BPS}
\end{figure}
The Fig.\ref{f_BPS} depicts the profiles of $f(z_{+})$ which show that $%
\gamma _{0}^{\left( n\right) }=f(0)<1$ for all $n$. The plots also show that 
$\gamma _{0}^{\left( n\right) }$ decreases when the values of $n$ increase: $%
\gamma _{0}^{\left( 1\right) }=0.398239039446192$\textbf{, }$\gamma
_{0}^{\left( 2\right) }=0.0823639322747160$\textbf{, }$\gamma _{0}^{\left(
3\right) }=0.00673809869063151$\textbf{, }$\gamma _{0}^{\left( 4\right)
}=0.000203456264553003$\textbf{, }$\gamma _{0}^{\left( 5\right)
}=0.00000225915820929528$, .... The numerical analysis shows that $\gamma
_{0}^{\left( n\right) }\rightarrow 0$ for large values of $n$. For all
values of $n$ and for large values of $z_{+}$ the profiles attain their
asymptotic value 1.

\begin{figure}[]
\centering\includegraphics[width=8.25cm]{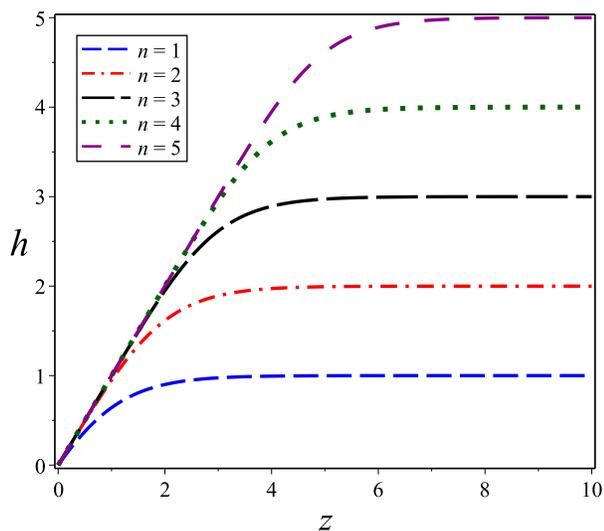}
\caption{The gauge field profile $f(z_+)$.}
\label{h_BPS}
\end{figure}
The Fig. \ref{h_BPS} shows the profiles of the gauge field $h(z_+)$. As
expected, they are null in $r=0$. The behavior, near the origin, for
sufficiently large values of $n$, is linear in $z_+$, in concordance with (%
\ref{bogo12}). For large values of $z_+$, the profiles reach their
asymptotic value $n$.

\begin{figure}[]
\centering\includegraphics[width=8.25cm]{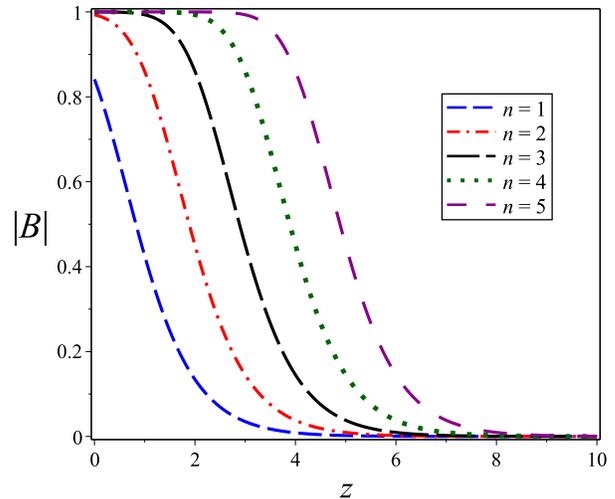}
\caption{The magnetic field $B(z_+)$.}
\label{b_BPS}
\end{figure}
The Fig. \ref{b_BPS} shows the profiles of the magnetic field $B(z_+)$. The
profiles, for low values of $n$ (here $n=1,2$), are lumps concentrated in
the origin, whose amplitude is $|B(0)|=1-(\gamma _{0})^{2}<1$. However, for
sufficiently large values of $n$ (here $n>3$), the profiles develop a
plateau whose width increases when $n$ increases. Such behavior resembles
that of the magnetic field of the Maxwell-Higgs vortices.

\begin{figure}[]
\centering\includegraphics[width=8.25cm]{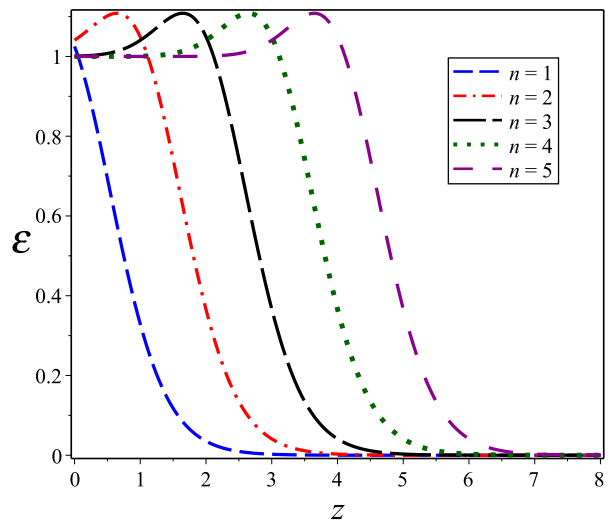}
\caption{The BPS energy density $\protect\varepsilon (z_+)$.}
\label{e_BPS}
\end{figure}

The Fig. \ref{e_BPS} describes the BPS energy density profiles $\varepsilon
(z_+)$. For $n=1$ is a lump centered at origin. For sufficiently large
values of $n$ (here $n>2$), the profiles form a plateau, starting at origin,
with amplitude 1 whose width increases when $n$ increases. At the ending of
the plateau the profile increases its values reaching a maximum value. After
this, the profiles decreases rapidly to zero. Such behavior also is shown in
the BPS energy density of the Maxwell-Higgs vortices.

In summary we study the abelian Maxwell-Higgs model on an infinite cylinder
surface. We explore the Bogomolnyi framework of the model in such surface,
showing that it is possible to construct a topological soliton solutions.
These solutions present interesting features, such as the behavior of the
field $\psi$ at $z_+=0$, which is different from zero, in contrast to
Nielsen-Olesen vortex solution \cite{NyO}, where $\phi(0)=0$. We also
analyze carefully the numerical solutions for the matter and gauge fields
and for the magnetic field and the energy density. Specifically, we have
analyzed solutions for $n>0$ however the solutions for $n<0$ (anti-soliton
solutions) can be easily visualized by doing: $h\rightarrow -h$, $%
f\rightarrow f$ and consequently $B\rightarrow -B$; it gives opposite
magnetic flux, as expected for an anti-soliton solution. It is worthwhile to
point out that there are also similar soliton solutions for $z<0$.

\begin{acknowledgments}
R.C. thanks to CNPq, CAPES and FAPEMA (Brazilian agencies) by financial
support. L.S. thanks Alan Kosteleck\'y and Gustavo Lozano for correspondence
and the Department of physics at Universidad de Buenos Aires for
hospitality. L.S. is supported by CONICET.
\end{acknowledgments}


\begin{thebibliography}{9}
\bibitem{NyO} H.B. Nielsen, P. Olesen, Nucl. Phys. B \textbf{61} 45 (1973).

\bibitem{hor} Peter A. Horvathy and Pengming Zhang,
Phys.Rept. \textbf{481}, 83 (2009).

\bibitem{hor1}
Peter A. Horvathy, [arXiv:hep-th/0704.3220].

\bibitem{hor2}
F. A. Schaposnik, [arXiv:hep-th/0611028].

\bibitem{hor3}
Gerald V. Dunne, [arXiv:hep-th/9902115].

\bibitem{WO} E. Witten and D. Olive, Phys. Lett. B \textbf{78} 97(1978).

\bibitem{Bogo} E. Bogomolyi, Sov. J. Nucl. Phys \textbf{24}, 449 (1976). 

\bibitem{Bogo1}
H. de Vega and F .A. Schaposnik, Phys. Rev. D \textbf{14}, 1100 (1976). 


\bibitem{thooft1}
Fabrizio Canfora, Gianni Tallarita,
[arXiv:hep-th/1407.0609], JHEP {\bf 1409}, 136 (2014).

\bibitem{hor0}
R.S. Ward, Nonlinearity {\bf 12}, 241 (1999).


\bibitem{hor01}
P. Sutcliffe, Phys.Rev. D {bf 85}, 125015 (2012).

\bibitem{hor02}
D. Harland, Phys.Lett. B {\bf 728}, 518 (2014).

\bibitem{hor03}
C. Adam, A. Wereszczynski, Phys.Rev. D {\bf 89}, 065010 (2014). 

\end{thebibliography}
\end{document}